\documentclass[10pt,conference]{IEEEtran}
\usepackage[utf8]{inputenc}
\usepackage[T1]{fontenc}
\usepackage{cite}

\usepackage{array}
\usepackage{booktabs}
\usepackage{svg}
\usepackage{graphicx}
\usepackage{amssymb}
\usepackage{amsmath}
\newcommand{\eqdef}{\overset{\cdot}{=}}

\usepackage{xcolor}

\DeclareMathOperator*{\argmin}{arg\,min\!}

\newcommand{\ma}[1]{\boldsymbol{#1}}
\newcommand{\compl}{\mathbb{C}}

\newcommand{\diagof}[1]{{\rm diag\!}\left\{ #1 \right\}}

\newcommand{\ten}[1]{\mathcal #1}
\newcommand{\fronorm}[1]{\left\|#1\right\|_{\rm F}}

\newcommand{\krp}{\diamond}
\newcommand{\kron}{\otimes}
\newcommand{\trans}{{{\rm T}}}
\newcommand{\herm}{{{\rm H}}}

\newcommand{\mat}[1]{\ensuremath{{\mathbf{#1}}}}
\newcommand{\vet}[1]{\mathit{\boldsymbol{#1}}}

\begin{document}
\title{Uplink Channel Estimation for Multi-User MISO Systems Assisted by a Fluid Reconfigurable Intelligent Surface}
\author{\IEEEauthorblockN{Amarilton L. Magalhães\IEEEauthorrefmark{1}\IEEEauthorrefmark{3}, André L. F. de Almeida\IEEEauthorrefmark{1}, and George C. Alexandropoulos\IEEEauthorrefmark{3}\IEEEauthorrefmark{4}}\\
\IEEEauthorblockA{Federal University of Ceará\IEEEauthorrefmark{1}, Federal Institute of Education, Science and Technology of Ceará\IEEEauthorrefmark{3}, Fortaleza, CE, Brazil\\
National and Kapodistrian University of Athens, Athens, Greece\IEEEauthorrefmark{3} / University of Illinois Chicago, IL, USA\IEEEauthorrefmark{4}\\E-mail: \{amarilton,andre\}@gtel.ufc.br, alexandg@di.uoa.gr}
}

\maketitle

\begin{abstract}
Fluid reconfigurable intelligent surfaces (FRISs) have recently emerged as a promising paradigm for wireless communications, wherein the reflecting elements can dynamically select their effective radiating positions from a dense preset grid, thereby introducing an additional degree of freedom. In contrast to conventional RIS architectures, FRISs can achieve spatial diversity with fewer physical elements. However, beyond the cascaded channel structure, FRIS-assisted systems are also affected by uncertainties arising from element-position mismatches caused by calibration inaccuracies or motion errors, which may degrade channel state information. To the best of our knowledge, channel estimation (CE) for FRIS-assisted systems under position uncertainty remains unexplored. To fill this gap, we propose a CE framework for a multi-user FRIS-assisted uplink system based on a two-time-scale FRIS configuration protocol that captures both reflection phase-shift and element-motion dynamics. By capitalizing on orthogonal pilot sequences and tensor modeling, we derive a closed-form solution that jointly estimates the individual channels and the motion-induced phase coefficients. Numerical results demonstrate notable performance in the presence of unknown position deviations.
\end{abstract}


\section{Introduction and motivation}\label{sec:introduction}
\IEEEPARstart{T}{he} growing interest in fluid antenna systems (FASs), an emerging paradigm of reconfigurable antennas that additionally dynamically adjusts antenna positions, offering a new degree of freedom for system design and signal processing, has recently attracted significant attention \cite{new2025atutorial}. In parallel, the widespread interest in reconfigurable intelligent surfaces (RISs) \cite{basar2024reconfigurable} has stimulated a closely related line of research referred to as fluid RIS (FRIS), which consists of a metasurface composed of response-tunable reflecting elements whose phase control port in each tile-type fluid element can be dynamically selected over a dense preset grid \cite{salem2026onthe,xiao2025from,ghadi2025performance2,xiao2025fluid}. Unlike conventional RISs, FRISs allow the physical positions of their constituent elements to vary over time, thereby enabling effective spatial diversity with even a single fluid element instead of a large number of fixed ones \cite{salem2025first}. In addition, recent work has proposed extending the FRIS coverage area by exploiting the principle of simultaneous transmitting and reflecting RIS (STAR-RIS), leading to the fluid integrated reflecting and emitting surface (FIRES) concept \cite{ghadi2025fires}.

To fully exploit the gains offered by FRISs, accurate channel state information (CSI) is essential. In classical RIS deployments, channel estimation (CE) can be impaired by hardware nonidealities and imperfect phase responses \cite{gomes2023channel,jian2022intelligent}. In FRIS-assisted systems, the CE task becomes even more challenging because, in addition to the cascaded bilinear channel structure, one must also account for position uncertainty arising from calibration and/or motion errors \cite{xiao2025from}, which may lead to mismatches between commanded and realized element positions. To the best of our knowledge, CE for FRIS-assisted systems has not yet been addressed in the literature. In this work, we propose a CE framework based on a FRIS-specific two-time-scale uplink training protocol for a multi-user FRIS-assisted communication system subject to element-position uncertainty. By exploiting the structure of the received signal and orthogonal pilot sequences, we derive an algebraic formulation that enables joint estimation of the individual channels and the motion-induced phase coefficients. We then adopt tensor-based modeling to obtain a closed-form solution. We compare the proposed CE method with a benchmark that assumes the realized motion-induced coefficients exactly match those commanded by the FRIS controller \cite{saggese2024onthe}. Simulation results demonstrate accurate estimation even in the presence of unknown motion deviations.

\textit{Notations}: $a$, $\vet{a}$, $\mat{A}$, and $\ten{A}$ denote scalars, vectors, matrices, and tensors, respectively. The operators $\mat{A}^\trans$ and $\mat{A}^\herm$ denote the transpose and Hermitian of $\mat{A}$. The Frobenius norm is denoted by $\fronorm{\cdot}$. The Hadamard, Khatri--Rao, Kronecker, and outer products are represented by $\odot$, $\krp$, $\kron$, and $\circ$, respectively. In addition, $\diagof{\vet{a}}$ denotes a diagonal matrix formed from $\vet{a}$.

\section{System model and assumptions}\label{sec:system}
\begin{figure*}[!t]
\minipage[t]{0.32\linewidth}
    \includegraphics[width=0.9\textwidth]{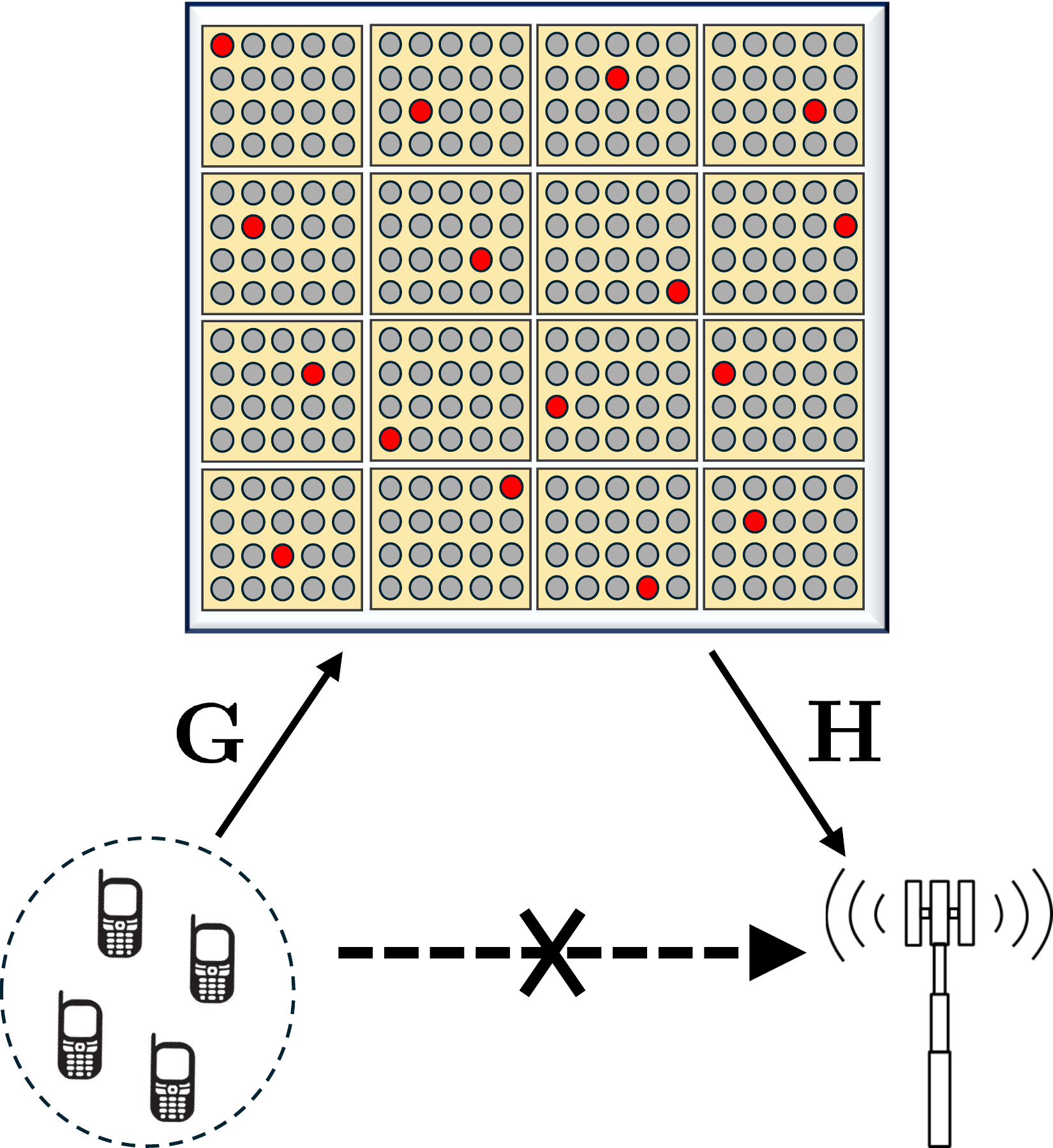}
    \caption{System sketch.}
    \label{fig:system}
\endminipage \hspace{-4ex}
\minipage[t]{0.32\linewidth}
    \includegraphics[width=\textwidth]{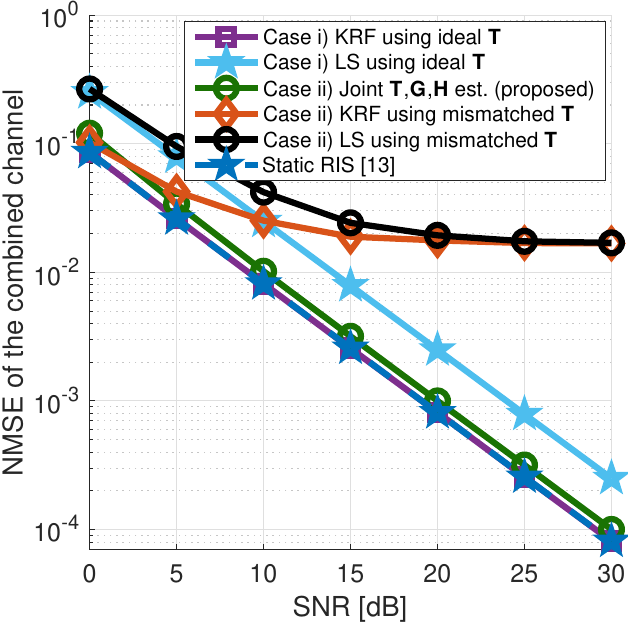}
    \caption{NMSE vs. SNR.}
    \label{fig:nmsesnr}
\endminipage ~\hspace{2ex}
\minipage[t]{0.32\linewidth}
    \includegraphics[width=\textwidth]{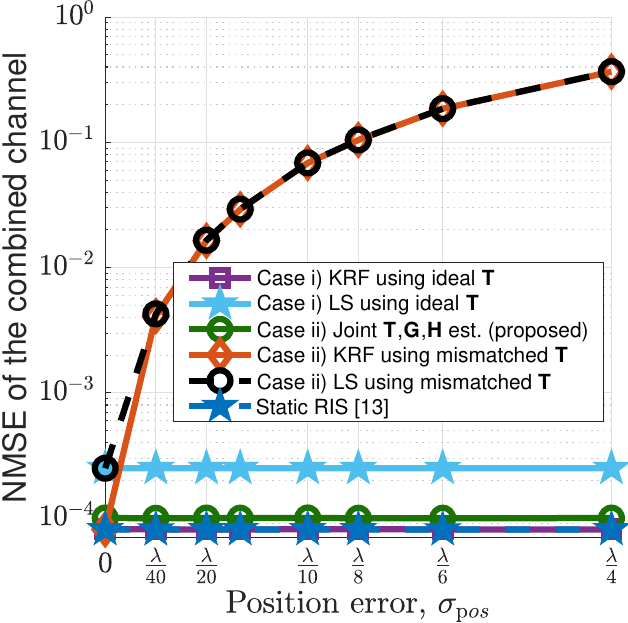}
    \caption{NMSE vs. position error.}
    \label{fig:nmsesigmapos}
\endminipage
\end{figure*}

Considering uplink transmission, we study a multi-user multiple-input single-output (MISO) wireless communication system in which $Q$ single-antenna user equipments (UEs) communicate with a base station (BS) equipped with $M_r$ antennas through an FRIS composed of $M$ fluid elements, as illustrated in Fig. \ref{fig:system}. The FRIS reflecting surface is assumed to occupy a total area $A$, and each $m$th ($m=1,\ldots,M$) fluid element (tile) can switch among $N$ preset positions. We consider a structured training window with $J$ sub-frames, each containing $K$ blocks of $T_s$ symbol periods, during which the propagation channels are assumed to remain constant. We adopt an FRIS reconfiguration strategy in which the electronic phase shift $e^{j\phi_{j,m}}$ remains fixed over the $K$ blocks of the $j$th sub-frame ($j=1,\ldots,J$) and varies across sub-frames, whereas the controlled reflecting positions $t_{k,m}$ vary across blocks and are repeated over the $J$ sub-frames. Specifically, $t_{k,m} = e^{-j\frac{2\pi}{\lambda}\|\vet{p}_{k,m}\|}$ is the motion-induced phase due to the $m$th fluid element's position at the $k$th block, given by the vector $\vet{p}_{k,m}$, with $\lambda$ denoting the wavelength. Defining $\mat{G} \in \compl^{M \times Q}$ and $\mat{H} \in \compl^{M_r \times M}$ as the UEs--FRIS and FRIS--BS channel matrices, respectively, the received signal after $T_s$ symbol periods in the $k$th block of the $j$th sub-frame is given by $\tilde{\mat{Y}}_{j,k} = \mat{H}\diagof{\vet{\phi}_j \odot \vet{t}_k}\mat{G}\mat{X}_p + \tilde{\mat{V}}_{j,k} \in \compl^{M_r \times T_s}$, where $\mat{X}_p \in \compl^{Q \times T_s}$ is the pilot matrix, $\vet{\phi}_j \eqdef [\phi_{j,1},\ldots,\phi_{j,M}]^\trans \in \compl^{M \times 1}$, $\vet{t}_k \eqdef [t_{k,1},\ldots,t_{k,M}]^\trans \in \compl^{M \times 1}$, and $\tilde{\mat{V}}_{j,k} \in \compl^{M_r \times T_s}$ is the associated noise term with zero-mean circularly symmetric complex-valued entries with variance $\sigma_v^2$. Assuming an orthogonal pilot design for $\mat{X}_p$, the matched-filtered signal at the BS is $\mat{Y}_{j,k} = (1/T_s)\tilde{\mat{Y}}_{j,k}\mat{X}_p^\herm \in \compl^{M_r \times Q}$, which can be written as
\begin{equation}\label{yjkfiltered}
    \mat{Y}_{j,k} = \mat{H}\diagof{\vet{\phi}_j \odot \vet{t}_k}\mat{G} + \mat{V}_{j,k},
\end{equation}
where $\mat{V}_{j,k} \in \compl^{M_r \times Q}$ is the filtered noise. Collecting all preset FRIS electro-mechanical configurations during the channel training phase, we define $\mat{T} \eqdef [\vet{t}_1,\ldots,\vet{t}_K]^\trans \in \compl^{K \times M}$ and $\ma{\Phi} \eqdef [\vet{\phi}_1,\ldots,\vet{\phi}_J]^\trans \in \compl^{J \times M}$ as the motion and phase-shift matrices, respectively. Equation \eqref{yjkfiltered} can be interpreted as the $(j,k)$th slice of a fourth-order tensor $\ten{Y} \in \compl^{M_r \times Q \times K \times J}$ obtained by concatenating $\mat{Y}_{j,k}$ along the third dimension for $k=1,\ldots,K$, and then along the fourth dimension for $j=1,\ldots,J$. This tensor admits a parallel factor (PARAFAC) decomposition \cite{kolda2009tensor,dearaujo2021channel,yuan2022channel} and can be written as $\ten{Y} = \ten{I}_{4,M} \times_1 \mat{H} \times_2 \mat{G}^\trans \times_3 \mat{T} \times_4 \ma{\Phi} + \ten{V}$, where $\ten{I}_{4,M}$ denotes the identity tensor and $\ten{V} \in \compl^{M_r \times Q \times K \times J}$ is the filtered noise tensor.

\section{Proposed channel estimation framework}
In this work, we consider two possible scenarios within the CE context. In the first scenario, we assume that the positions of the FRIS elements exactly match those commanded by the controller. In this case, our goal is to estimate the individual channels $\mat{G}$ and $\mat{H}$ with the knowledge of $\ma{\Phi}$ and $\mat{T}$. To this, we consider solving the problem
\begin{equation}\label{opt1}
    \min \limits_{\mat{G},\mat{H}} \Big\|\ten{Y} - \ten{I}_{4,M} \times_1 \mat{H} \times_2 \mat{G}^\trans \times_3 \mat{T} \times_4 \ma{\Phi}\Big\|^2_{\textrm{F}}.
\end{equation}
Exploiting the multilinear structure of $\ten{Y}$, we unfold it into the matrix $\mat{Y}_1 = \ma{\Theta}\bigl(\ma{\Phi} \krp \mat{T}\bigr) + \mat{V}_1 \in \compl^{QM_r \times JK}$ (see \cite{kolda2009tensor} for details), where $\ma{\Theta} \eqdef \mat{G}^\trans \krp \mat{H} \in \compl^{QM_r \times M}$ is the combined channel, with $\mat{V}_1$ denoting the corresponding noise term. An estimate of $\ma{\Theta}$ can be obtained by solving the least-squares (LS) problem $\hat{\ma{\Theta}} = \argmin_{\ma{\Theta}} \bigl\|\mat{Y}_1 -  \ma{\Theta}\bigl(\ma{\Phi} \krp \mat{T}\bigr)\bigr\|_\mathrm{F}^2$. Assuming a semi-unitary design for $\ma{\Phi}$, this estimate is given by the linear filter $\hat{\ma{\Theta}} = (1/JK)\mat{Y}\bigl(\ma{\Phi} \krp \mat{T}\bigr)^\ast$. To provide decoupled estimates of $\mat{G}$ and $\mat{H}$, we consider the problem
\begin{equation}\label{minkrf}
    \min \limits_{\mat{G},\mat{H}} \Big\|\hat{\ma{\Theta}} - \mat{G}^\trans \krp \mat{H}\Big\|^2_{\textrm{F}}.
\end{equation}
This problem can be tackled by invoking the Khatri-Rao factorization (KRF) algorithm \cite{Kibangou2009}, which involves solving $M$ parallelized rank-1 approximations, explored in the literature for a static RIS with fixed elements \cite{dearaujo2021channel}.

In the second scenario, the commanded positions are not perfectly achieved due to motion errors, resulting in imprecise CSI. To account for this, it is crucial to explicitly estimate the motion matrix $\mat{T}$ for accurate CE. For this purpose, we propose to include the matrix $\mat{T}$ in the minimization problem \eqref{opt1}, yielding a generalized problem to solve three quantities in place of two, namely, $\mat{T}$, $\mat{G}$, and $\mat{H}$. To tackle it, we now unfold $\ten{Y}$ into $\mat{Y}_2 = \mat{Z}\ma{\Phi} + \mat{V}_2 \in \compl^{KQM_r \times J}$, where $\mat{Z} \eqdef \mat{T} \krp \mat{G}^\trans \krp \mat{H} \in \compl^{KQM_r \times M}$ captures the joint effect of the motion and the cascaded channel, while $\mat{V}_2$ represents the corresponding noise term. Likewise, we initially can solve $\hat{\mat{Z}} = \argmin_{\mat{Z}} \bigl\|\mat{Y} - \mat{Z}\ma{\Phi}\bigr\|_\mathrm{F}^2$, for which the analytical solution is $\hat{\mat{Z}} = (1/J)\mat{Y}\ma{\Phi}^\ast$. We then aim to decouple the three factors from $\hat{\mat{Z}}$ by considering the following problem:
\begin{equation}\label{minkrf3}
    \min \limits_{\mat{T},\mat{G},\mat{H}} \Big\|\hat{\mat{Z}} - \mat{T} \krp \mat{G}^\trans \krp \mat{H}\Big\|^2_{\textrm{F}}.
\end{equation}
We address problem \eqref{minkrf3} by employing multidimensional LS Khatri--Rao factorization \cite{weis2012dual}, which can be viewed as a generalization of \eqref{minkrf}. We approximate the $m$th column of $\hat{\mat{Z}}$ as $\vet{z}_m \eqdef \vet{z}_m^{(\mat{T})} \kron \vet{z}_m^{(\mat{G})} \kron \vet{z}_m^{(\mat{H})} \in \compl^{KQM_r \times 1}$, where $\vet{z}_m^{(\mat{T})}$, $\vet{z}_m^{(\mat{G})}$, and $\vet{z}_m^{(\mat{H})}$ denote the $m$th columns of $\mat{T}$, $\mat{G}^\trans$, and $\mat{H}$, respectively. Tensorizing $\vet{z}_m$ yields the rank-one tensor $\ten{Z}_r = \vet{z}_m^{(\mat{H})} \circ \vet{z}_m^{(\mat{G})} \circ \vet{z}_m^{(\mat{T})} \in \compl^{M_r \times Q \times K}$. The columns of $\hat{\mat{T}}$, $\hat{\mat{G}}^\trans$, and $\hat{\mat{H}}$ are then obtained via a rank-one truncated higher-order singular value decomposition (HOSVD) \cite{de2000multilinear}, namely, $\vet{z}_m^{(\mat{T})} = \sqrt[3]{(\ten{S}_r)_{1,1,1}}\vet{u}^{(3)}_{m,1}$, $\vet{z}_m^{(\mat{G})} = \sqrt[3]{(\ten{S}_r)_{1,1,1}}\vet{u}^{(2)}_{m,1}$, and $\vet{z}_m^{(\mat{H})} = \sqrt[3]{(\ten{S}_r)_{1,1,1}}\vet{u}^{(1)}_{m,1}$, where $\ten{S}_r$ is the core tensor and $\vet{u}^{(1)}_{m,1}$, $\vet{u}^{(2)}_{m,1}$, and $\vet{u}^{(3)}_{m,1}$ are the dominant higher-order singular vectors of $\ten{Z}_m$ along modes 1, 2, and 3, respectively. Repeating this procedure for $m=1,\ldots,M$ yields the estimates of $\mat{T}$, $\mat{G}$, and $\mat{H}$.

\begin{table*}[!t]
\centering
\caption{Comparison of FRIS motion knowledge regimes}
\label{tab:comparison_motion}
\begin{tabular}{>{\centering\arraybackslash}m{3cm} >{\centering\arraybackslash}m{2cm} >{\centering\arraybackslash}m{2cm} >{\centering\arraybackslash}m{2cm} >{\centering\arraybackslash}m{2cm} >{\centering\arraybackslash}m{3cm}}
        \midrule
        Scenario & Motion model & Unknown factors & Physical realism& Identifiability & Complexity analysis\\
        \midrule
        \textbf{Perfect Motion}&$\mat{T}$ known&$\mat{G}$, $\mat{H}$&Idealized&\\
        \textbf{Imperfect Motion}&$\mat{T}$ unknown&$\mat{T}$, $\mat{G}$, $\mat{H}$&Conservative&\multicolumn{2}{c}{\textit{To be included in the full version}}\\
        \textbf{Calibrated Motion}$^{(\ast)}$&  $\mat{T}=\mat{T}_{(0)} + \ma{\Delta}_{\mat{T}}$& $\ma{\Delta}_{\mat{T}}$, $\mat{G}$, $\mat{H}$ & Realistic & \\
        \midrule
	\end{tabular}\\
$^{(\ast)}$Further details on the calibrated motion model and its impact on the proposed channel estimation approach will be addressed in the full version.
\end{table*}

\section{Performance evaluation and upcoming results}\label{sec:results}
We present preliminary simulation results to evaluate the performance of the proposed CE methods. In particular, we report the normalized mean square error (NMSE) for two motion cases: i) perfect motion realization, in which the realized positions coincide with the commanded ones; and ii) motion realization uncertainty, where mismatches arise due to calibration and/or motion errors. In the former, the knowledge of the FRIS-commanded motion matrix $\mat{T}$ can be leveraged to get the channel estimates, whereas in the latter, it is preferable to treat $\mat{T}$ as unknown (blind motion) and jointly estimate it. We adopt the parameter set $\{M,Q,M_r,T_s,K\} = \{12,4,10,4,4\}$, and we design the pilot matrix $\mat{X}_p$ as a truncated-discrete Fourier matrix. We first provide NMSE results for CE in Fig. \ref{fig:nmsesnr} by ranging the signal-to-noise ratio (SNR). Accounting for case i), (refer to \textit{`KRF using ideal $\boldsymbol{T}$'} and \textit{`LS using ideal $\boldsymbol{T}$'} curves), KRF results approach the performance of that of classical RIS with fixed elements \cite{dearaujo2021channel}, in contrast to the LS methods. In the algebraic sense, the methods in case i) employ similar filtering as \cite{dearaujo2021channel} after collecting $JK$ signal blocks. However, $\hat{\ma{\Theta}}$ is estimated more accurately employing KRF than that of the conventional LS method. This improvement arises from decoupling the individual channel components using KRF, leveraging its denoising properties, and then recombining them. This gain was also reported by \cite{dearaujo2021channel} in the context of static RIS. To assess the impact of motion errors on CE performance, we now evaluate case ii). For this purpose, we define a wavelength-dependent positioning error $\sigma_{\mathrm{pos}}$ which directly affects the motion matrix$\mat{T}$, and set $\sigma_{\mathrm{pos}} = \lambda/20$. We initially focus on the receiver processing in which the BS ignores the motion errors (refer to \textit{`KRF using mismatched $\boldsymbol{T}$'} and \textit{`LS using mismatched $\boldsymbol{T}$'} curves). The worst performance of the LS method for the low to medium SNR regimes is due to not refining the CE, similar to case i), unlike KRF one. Nevertheless, the KRF's denoising gain is no longer effective when motion errors are present, where the commanded $\mat{T}$ does not coincide with the realized. Still in Fig. \ref{fig:nmsesnr}, our proposed method (refer to \textit{`Joint \it $\boldsymbol{T}$, $\boldsymbol{G}$, $\boldsymbol{H}$ estimation (proposed)'} curve) jointly estimates the motion matrix $\mat{T}$ and the individual channel components under motion uncertainty, thereby tackling the previous limitation and achieving substantial performance enhancement, approaching to those of case i). In Fig. \ref{fig:nmsesigmapos}, we present a comparison by taking a closer look at the motion error. To this end, we set a wavelength-dependent positioning error $\sigma_{\mathrm{pos}}$ while adopting SNR = 30 dB. Results in Fig. \ref{fig:nmsesigmapos} corroborate the behavior previously observed, emphasizing the need to account for position uncertainties. The proposed method remains robust to increasing error levels, approaching the performance of the static RIS and the idealized-motion FRIS (benchmark). This does not hold for the scheme that neglects the estimation of the motion matrix $\mat{T}$, which suffers from pronounced performance degradation. Table \ref{tab:comparison_motion} summarizes the scenarios considered in this work.

\

\textit{Next steps}: The full paper will include analyzing the impact of the number of FRIS elements and users on system performance. A more detailed description of the proposed algorithms, the formulation of alternative iterative channel estimation methods, a discussion on identifiability conditions, and computational complexity analyses will also be provided. In addition, a calibrated-motion approach and a more extensive simulation campaign will be incorporated.

\bibliographystyle{IEEEtran}
\bibliography{references}

@article{new2025atutorial,
  author={New, Wee Kiat and Wong, Kai-Kit and Xu, Hao and Wang, Chao and Ghadi, Farshad Rostami and Zhang, Jichen and Rao, Junhui and Murch, Ross and Ramírez-Espinosa, Pablo and Morales-Jimenez, David and Chae, Chan-Byoung and Tong, Kin-Fai},
  journal={IEEE Communications Surveys \& Tutorials}, 
  title={A Tutorial on Fluid Antenna System for 6{G} Networks: Encompassing Communication Theory, Optimization Methods and Hardware Designs}, 
  year={2025},
  volume={27},
  number={4},
  pages={2325-2377}
  }

@article{salem2025first,
  title={A first look at the performance enhancement potential of fluid reconfigurable intelligent surface},
  author={Salem, Abdelhamid and Wong, Kai-Kit and Alexandropoulos, George and Chae, Chan-Byoung and Murch, Ross},
  journal={arXiv preprint arXiv:2502.17116},
  year={2025}
}

@article{ghadi2025fires,
  author={Ghadi, Farshad Rostami and Wong, Kai-Kit and Kaveh, Masoud and López-Martínez, F. Javier and Chae, Chan-Byoung and Alexandropoulos, George C.},
  journal={IEEE Wireless Communications Letters}, 
  title={{FIRES}: Fluid Integrated Reflecting and Emitting Surfaces}, 
  year={2025},
  volume={14},
  number={11},
  pages={3744-3748}
}

@inproceedings{Kibangou2009,
    title={Non-iterative solution for {PARAFAC} with a {T}oeplitz matrix factor},
    author={A. Y. {Kibangou} and G. {Favier}},
    booktitle={Proc. EUSIPCO},
    pages={691-695},
    month={8},
    year={2009},
}

@article{salem2026onthe,
  title={On the performance enhancement potential of fluid reconfigurable intelligent surfaces},
  author={Salem, Abdelhamid and Wong, Kai-Kit and Alexandropoulos, George and Chae, Chan-Byoung and Murch, Ross},
  journal={IEEE Transactions on Wireless Communications},
  year={2026}
}

@article{xiao2025from,
  title={From fixed to fluid: Unlocking the new potential with fluid {RIS} ({FRIS})},
  author={Xiao, Han and Hu, Xiaoyan and Wong, Kai-Kit and Zhu, Xusheng and Hong, Hanjiang and Ghadi, Farshad Rostami and Xu, Hao and Chae, Chan-Byoung},
  journal={arXiv preprint arXiv:2509.18899},
  year={2025}
}

@article{gomes2023channel,
	title={Channel estimation in {RIS}-assisted {MIMO} systems operating under imperfections},
	author={Gomes, Paulo R B and de Ara{\'u}jo, Gilderlan T and Sokal, Bruno and de Almeida, Andr{\'e} L F and Makki, Behrooz and Fodor, G{\'a}bor},
	journal={IEEE Trans. Veh. Technol.},
	year={2023},
	publisher={IEEE}
}

@inproceedings{weis2012dual,
  title={Dual-symmetric parallel factor analysis using procrustes estimation and {K}hatri-{R}ao factorization},
  author={Weis, Martin and Roemer, Florian and Haardt, Martin and Husar, Peter},
  booktitle={2012 Proceedings of the 20th European Signal Processing Conference (EUSIPCO)},
  pages={270--274},
  year={2012},
  organization={IEEE}
}

@article{de2000multilinear,
  title={A multilinear singular value decomposition},
  author={De Lathauwer, Lieven and De Moor, Bart and Vandewalle, Joos},
  journal={SIAM Journal on Matrix Analysis and Applications},
  volume={21},
  number={4},
  pages={1253--1278},
  year={2000},
  publisher={SIAM}
}

@article{dearaujo2021channel,
	title={Channel estimation for intelligent reflecting surface assisted {MIMO} systems: A tensor modeling approach},
	author={de Ara{\'u}jo, Gilderlan T and de Almeida, Andr{\'e} L F and Boyer, R{\'e}my},
	journal={IEEE J. Sel. Topics Signal Process.},
	volume={15},
	number={3},
	pages={789--802},
	year={2021},
	publisher={IEEE}
}

@article{kolda2009tensor,
	title={Tensor decompositions and applications},
	author={Kolda, Tamara G and Bader, Brett W},
	journal={SIAM Review},
	volume={51},
	number={3},
	pages={455--500},
	year={2009},
	publisher={SIAM}
}

@article{basar2024reconfigurable,
  author={Basar, Ertugrul and Alexandropoulos, George C. and Liu, Yuanwei and Wu, Qingqing and Jin, Shi and Yuen, Chau and Dobre, Octavia A. and Schober, Robert},
  journal={IEEE Vehicular Technology Magazine}, 
  title={Reconfigurable Intelligent Surfaces for 6{G}: Emerging Hardware Architectures, Applications, and Open Challenges}, 
  year={2024},
  volume={19},
  number={3},
  pages={27-47}
}

@article{ghadi2025performance2,
  author={Rostami Ghadi, Farshad and Wong, Kai-Kit and Javier López-Martínez, F. and Alexandropoulos, George C. and Chae, Chan-Byoung},
  journal={IEEE Wireless Communications Letters}, 
  title={Performance Analysis of Wireless Communication Systems Assisted by Fluid Reconfigurable Intelligent Surfaces}, 
  year={2025},
  volume={14},
  number={12},
  pages={3922-3926}
}

@article{xiao2025fluid,
  author={Xiao, Han and Hu, Xiaoyan and Wong, Kai-Kit and Hong, Hanjiang and Alexandropoulos, George C. and Chae, Chan-Byoung},
  journal={IEEE Wireless Communications Letters}, 
  title={Fluid Reconfigurable Intelligent Surfaces: Joint {O}n-{O}ff Selection and Beamforming With Discrete Phase Shifts}, 
  year={2025},
  volume={14},
  number={10},
  pages={3124-3128}
}

@article{saggese2024onthe,
  author={Saggese, Fabio and Croisfelt, Victor and Kotaba, Radosław and Stylianopoulos, Kyriakos and Alexandropoulos, George C. and Popovski, Petar},
  journal={IEEE Open Journal of the Communications Society}, 
  title={On the Impact of Control Signaling in {RIS}-Empowered Wireless Communications}, 
  year={2024},
  volume={5},
  number={},
  pages={4383-4399}
}

@article{jian2022intelligent,
  author={Jian, Mengnan and Alexandropoulos, George C. and Basar, Ertugrul and Huang, Chongwen and Liu, Ruiqi and Liu, Yuanwei and Yuen, Chau},
  journal={Intelligent and Converged Networks}, 
  title={Reconfigurable intelligent surfaces for wireless communications: Overview of hardware designs, channel models, and estimation techniques}, 
  year={2022},
  volume={3},
  number={1},
  pages={1-32}
}

@inproceedings{yuan2022channel,
  author={Yuan, Jide and Alexandropoulos, George C. and Kofidis, Eleftherios and Jensen, Tobias Lindstrom and De Carvalho, Elisabeth},
  booktitle={2022 IEEE International Conference on Communications Workshops (ICC Workshops)}, 
  title={Channel Tracking for {RIS}-Enabled Multi-User {SIMO} Systems in Time-Varying Wireless Channels}, 
  year={2022},
  volume={},
  number={},
  pages={145-150}
}

\end{document}